\documentclass[aps,prb,twocolumn,superscriptaddress,showpacs,floatfix,amsmath,amssymb]{revtex4-1}
\usepackage{amssymb}
\usepackage{bm}
\usepackage{url}
\usepackage{graphicx}
\usepackage{color}
\begin{document}

\title{Gyrotactic trapping in laminar and turbulent Kolmogorov flow} 

\author{Francesco Santamaria}
\affiliation{Dipartimento di Fisica and INFN, Universit\`a di Torino, 
via P. Giuria 1, 10125 Torino, Italy}

\author{Filippo De Lillo}
\affiliation{Dipartimento di Fisica and INFN, Universit\`a di Torino, 
via P. Giuria 1, 10125 Torino, Italy}

\author{Massimo Cencini}
\affiliation{Istituto dei Sistemi Complessi, CNR, via dei Taurini 19, 00185 Rome, Italy}

\author{Guido Boffetta}
\affiliation{Dipartimento di Fisica and INFN, Universit\`a di Torino, 
via P. Giuria 1, 10125 Torino, Italy}

\begin{abstract}
Phytoplankton patchiness, namely the heterogeneous distribution of microalgae over multiple spatial scales, dramatically impacts marine ecology.  A spectacular example of such heterogeneity occurs in thin phytoplankton layers (TPLs), where large numbers of photosynthetic microorganisms are found within a small depth interval.  Some species of motile phytoplankton can form TPLs by gyrotactic trapping due to the interplay of their particular swimming style (directed motion biased against gravity) and the transport by a flow with shear along the direction of gravity.  Here we consider gyrotactic swimmers in numerical simulations of the Kolmogorov shear flow, both in laminar and turbulent regimes.  In the laminar case, we show that the swimmer motion is integrable and the formation of TPLs can be fully characterized by means of dynamical systems tools. We then study the effects of rotational Brownian motion or turbulent fluctuations (appearing when the Reynolds number is large enough) on TPLs. In both cases we show that TPLs become transient, and we characterize their persistence.
\end{abstract}


\maketitle

\section{Introduction\label{sec:1}}

Motile aquatic microorganisms in their natural habitats move under the
simultaneous and combined effect of ambient transport (currents,
turbulence etc.) and swimming.\cite{Pedley1987,Pedley1992} The
interaction between these different transport mechanisms can give rise to
interesting phenomena, such as the generation of
inhomogeneous distributions\cite{Kessler1985,Durham2009} and
swimming-induced flows like bioconvection\cite{PedleyHill1988,williams2011} or bacterial turbulence,\cite{Dunkel2013} many aspects of which can be
studied within the theoretical framework of dynamical systems theory
and fluid mechanics.\cite{Torney2007,Thorn2010,Stark_PRL2012,Stark_EPJE2013,chacon2013,Durham2013,DeLillo2014}

For most microorganisms, swimming is biased in specific directions by
some kind of {\it taxis} in response to chemical
(e.g. \textit{chemotaxis}\cite{Berg2004}) or physical signals
(e.g. \textit{phototaxis}\cite{Garcia2013} and
\textit{magnetotaxis}\cite{magnetotaxis}). One of those taxes,
relevant to several species of phytoplankton, tends to orient cell
swimming direction upward against gravity (negative
\textit{gravitaxis}). Although other mechanisms are also possible,
vertical orientation typically results from the gravitational torque
due to the asymmetric cell-density distribution, leading to bottom
heaviness.\cite{Kessler1985,Pedley1987,Pedley1992} In the presence of
a flow, gravitational torque combines with the hydrodynamic one giving
rise to directed locomotion, dubbed {\it gyrotaxis}, which can
eventually cause accumulation of cells in specific flow regions.  In a
laminar downwelling pipe flow, for instance, the interplay of swimming
and hydrodynamic shear produces a striking aggregation in the center
of the pipe known as gyrotactic focusing.\cite{Kessler1985} In
homogeneous isotropic turbulence, numerical simulations have shown
that gyrotactic algae generate small-scale clusters with fractal
distributions.\cite{Durham2013,DeLillo2014} Such findings, which have
been rationalized using tools from dynamical systems and fluid
mechanics, may be an explanation of why field observations have found
that small-scale patchiness appears to be stronger in motile
phytoplankton species.\cite{malkiel1999,gallager2004,mouritsen2003}

In this paper we consider a case intermediate between laminar flows
and homogeneous turbulence: turbulence in the presence of a mean shear
flow. The motivation for our study comes from the recent experimental
observation that gyrotactic algae swimming within a laminar vertical
shear aggregate in horizontal layers around the maximal shear rate, as
a consequence of \textit{gyrotactic trapping}.\cite{Durham2009} This
mechanism has been proposed as a possible explanation, at least for
some phytoplankton species, for the formation of the spectacular thin
phytoplankton layers (TPLs) often observed in (coastal) ocean.  TPLs
are high concentrations of phytoplankton, centimeters to one meter
thick, which extend horizontally up to kilometers and last from hours
to a few
days.\cite{Dekshenieks2001,cmvgdc_ec07,cd_ms09,Steinbuck2009,Durham2012}
They are important to marine ecology by enhancing zooplankton growth
rates, thus providing high concentration of preys for fishes and
their larvae.  Moreover, as many phytoplankton
species found in TPLs are toxic, their presence can enhance
zooplankton and fish mortality, or induce zooplankton to avoid toxic
and mucus rich layers.  TPLs can be formed by several motile and
non-motile species, therefore very likely there is not a unique
mechanism for their formation. See the review
Ref.\onlinecite{Durham2012} and references therein for an up-to-date
account on various aspects of TPLs.

It is worth recalling the basic ideas of the microfluidic
experiments,\cite{Durham2009} which have demonstrated gyrotactic
trapping for {\it Chlamydomonas nivalis} and {\it Heterosigma
  akashiwo} (a toxic species). Swimming algae were injected at the
bottom of a centimeter-sized tank where a vertical shear is induced by
a rotating belt. Cells swim upwards to about the middle of the tank
where the shear rate becomes sufficiently strong to overcome
gravitational bias and cause the swimming direction to tumble. Loosing
gravitational bias, no net vertical velocity can be maintained. Hence,
cells remain trapped, accumulating in horizontal layers. Recent
numerical simulations have shown that gyrotactic phytoplankton forms
thin layers even in non-stationary Kelvin-Helmholtz flow, where
swimming cells are found to be trapped in evolving KH
billows.\cite{hs_csr12} However, field experiments with simultaneous
measurement of biological and physical properties have shown that
while thin layers are weakly affected by turbulence of moderate
intensity, stronger turbulence will dissolve
them.\cite{wg_csr10,Sullivan2010} The entire process is rather
nontrivial and poorly understood:\cite{Durham2012} on the one hand
shear flows can induce layers by gyrotactic trapping; on the other
hand they trigger the generation of turbulence which, in turn, can
destabilize gyrotactic trapping causing layers' break-up.

In the present paper we study phytoplankton layers in the Kolmogorov
flow with shear along the vertical direction. This is a well known
periodic shear flow model for studying the transition to
turbulence\cite{Sivashinsky1985,She1987,Borue1996} and it is presented together
with the model equations for gyrotactic motion in Sect.~\ref{sec:2}.

In laminar and steady Kolmogorov flow (Sect.~\ref{sec:3}), we have
been able to  solve gyrotactic swimmer dynamics by using
tools from dynamical systems. In particular, we have found that the
motion is integrable, allowing us to analytically characterize
gyrotactic trapping. We have then numerically studied
the effect of stochasticity, namely rotational Brownian motion, on the
evolution of the thin layers (Sect.~\ref{sec:4}). 

In the turbulent Kolmogorov flow, small-scale fluctuations superimpose
to the mean large scale flow, and no analytical study is possible.
Direct numerical simulations of the Navier-Stokes equations, coupled
with the Lagrangian dynamics of gyrotactic swimmers, have been used to
investigate the dynamical effects of turbulent fluctuations on TPLs
(Sect.~\ref{sec:5}).

Both stochastic effects and turbulence make TPL a transient
phenomenon, and we characterized its persistence
properties. Discussions and final remarks on the relevance of our
findings are presented in Sect.~\ref{sec:6}.

\section{Models\label{sec:2}}

\subsection{Gyrotactic swimming \label{sec:2.1}}
We consider spherical cells, which is justified by detailed analysis
of cell morphology,\cite{OMalley2012} and dilute suspensions so that
alga-alga interactions can be neglected as well as back-reaction on
the fluid flow. Moreover, thanks to the small size of the cells ($\sim
10\;\mu m$) with respect to the Kolmogorov length scale ($\eta$,
namely the smallest scale of turbulent flows, in oceans $\eta\sim
0.3-10 \;mm$) they can be considered as point particles and 
their motion is akin to that of  passive tracers but for their 
ability to swim.
We assume that cells
are neutrally buoyant, as their sedimentation speed ($\sim 2.5-3\;\mu
m/s$) is much smaller than their typical swimming speed ($\sim 100 \;
\mu m/s$).\cite{PedleyHill1988,OMalley2012}

According to the classic model of gyrotactic motility\cite{Kessler1985,Pedley1987,Pedley1992} the
position $\bm X$ and the swimming orientation ${\bf p}$ (where $|{\bf
  p}|=1$) of a gyrotactic cell evolve according to the equations
\begin{eqnarray}
\dot{\bm X} &=& {\bm u} + v_s {\bf p}\,,
\label{eq:1}\\
\dot{\bf p} &=& \frac{1}{2B}\left[\hat{\bm z}  - (\hat{\bm z} \cdot {\bf p}) {\bf p}\right] +\frac{1}{2} {\bm \omega} \times {\bf p}\,,
\label{eq:2}
\end{eqnarray}
$\hat{\bm z}$ denoting the vertical unit vector.  In Eq.~(\ref{eq:1}),
the cell velocity is given by the superposition of the fluid velocity
at the cell location, $\bm u({\bm X},t)$, and the swimming velocity,
$v_s {\bf p}$, with $v_s$ assumed to be
constant.\cite{Pedley1987,Pedley1992} As for the swimming direction
dynamics, the first term on the r.h.s. of Eq.~(\ref{eq:2}) accounts
for the bias in the direction opposite to gravitational acceleration,
$\bm g=-g \hat{\bm z}$, with a characteristic orientation time $B$ (in
a still fluid, $\bm u=0$, $B$ is the typical time a cell employs to
orient upwards). For bottom-heavy, neutrally buoyant and spherical
cells in a fluid with kinematic viscosity $\nu$ we have $B=3
\nu/(hg)$, $h$ measuring the distance between the cell center of mass
and its geometric center.  We remark that this term has, in general,
an additional contribution arising from fluid
acceleration.\cite{DeLillo2014} However for the formation of TPLs in
the oceans, where turbulence is not very intense, with typical values
of the turbulent energy dissipation $\epsilon \ll 10^{-4}m/s^3$ (see
e.g. Ref.\onlinecite{Thorpe2007}), fluid acceleration ($\sim
(\epsilon^3/\nu)^{1/4}\approx 0.1 \,m/s^2 \ll g$) can be safely
neglected.  Finally, the last term in (\ref{eq:2}) represents the
rotation of the swimming direction due to fluid vorticity ${\bm
  \omega}={\bm \nabla} \times {\bm u}$.
 
By comparing the two terms in Eq.~(\ref{eq:1}), we can define the {\it
  swimming number} $\Phi=v_S/U$ where $U$ is a typical velocity of the
flow, providing a dimensionless measure of the swimming velocity.
While from Eq.~(\ref{eq:2}) we obtain the dimensionless {\it stability
  number} $\Psi=B\omega$, where $\omega$ is a measure of the typical
vorticity intensity.  The latter number measures the importance of
vortical overturning with respect to directional
swimming.\cite{Durham2013} Given the flow, specified in the following
Section, the values of these two numbers determine the behavior of the
swimming cells.

In Sect.~\ref{sec:4} we will also consider the presence of stochastic
terms (rotational Brownian motion) in Eq.~(\ref{eq:2}).

\subsection{The Kolmogorov flow\label{sec:2.2}}

As discussed in the Introduction, gyrotactic cells in vertical shears
can form thin layers when shear vorticity exceeds the inverse
orientation time, i.e. when $\Psi>1$, as demonstrated in laboratory
experiments.\cite{Durham2009} However, shear-induced turbulence can
dissolve the layers after a finite lifetime. The process of TPL
break-up due to turbulent fluctuations and thus the persistence
properties of TPLs are still poorly characterized, mainly because of
the experimental difficulties in tracking TPLs from birth to
death.\cite{Durham2012} Aiming to numerically explore the effects of
turbulent fluctuations on the gyrotactic trapping, we consider here
the periodic shear flow, originally introduced by Kolmogorov to study
the transition to turbulence. Several analytical studies have
investigated its linear stability properties and weakly nonlinear
behavior.\cite{Sivashinsky1985} Moreover, extensive numerical
simulations have explored the fully turbulent
regime.\cite{Borue1996,Musacchio2014}

The Kolmogorov flow is realized when the
  Navier-Stokes equation for an incompressible fluid ($\bm \nabla
  \cdot \bm u=0$), is sustained via the Kolmogorov body force, i.e.
\begin{equation}
\partial_t \bm u +\bm u \cdot \bm \nabla \bm u = - \bm \nabla p +\nu \Delta \bm u + F\cos(z/L) \hat{\bm x}\, 
\label{eq:ns}
\end{equation}
where $p$ is the pressure, density is taken to unity $\rho=1$, and
$\hat{\bm x}$ denotes the unit vector in the horizontal direction.
The physical domain is a cube of size $L_B=2\pi L$ with periodic
boundary conditions in all directions.  It is easy to verify that
(\ref{eq:ns}) admits a stationary solution, the {\it laminar}
Kolmogorov flow $\bm u= U \cos(z/L) \hat{\bm x}$ with $U= L^2 F/\nu$.
This laminar solution becomes unstable with respect to transverse
perturbations on scales larger than $L$ when the Reynolds number,
$Re=UL/\nu$, exceeds the critical value
$Re_c=\sqrt{2}$.  The first instability is two-dimensional (thanks to
the Squire's theorem, valid for 2D, parallel flows) but, by increasing
$Re$, three-dimensional motion develops and the flow eventually
becomes turbulent.\cite{Borue1996,Musacchio2014} Remarkably, even in
the fully developed turbulent state the mean velocity profile
$\bar{\bm u}$ (the over-bar denoting time average) remains
monochromatic, as in the laminar flow, i.e. $\bar{\bm u}=U \cos(z/L)
\hat{\bm x}$, with a different amplitude $U<L^2
F/\nu$.\cite{Musacchio2014} By changing the relative amplitude of the
turbulent fluctuations with respect to the mean Kolmogorov flow, we
will investigate the effects of turbulence on shear-induced gyrotactic
trapping, and thus on the persistence and properties of the resulting
thin layers.

\section{Swimming in the laminar Kolmogorov flow \label{sec:3}}

We start considering gyrotactic microorganisms swimming in a laminar
Kolmogorov flow $\bm u= U \cos(z/L) \hat{\bm x}$. It is
useful to make Eqs.~(\ref{eq:1}) and (\ref{eq:2}) non-dimensional by
measuring lengths, velocities and times in terms of $L$, $U$ and
$L/U$. In particular, Eq.~(\ref{eq:1}) reads
\begin{eqnarray}
\dot{X} &=& \cos{Z} + \Phi \mathrm{p}_x \label{eq:xdot}\\
\dot{Y} &=& \Phi \mathrm{p}_y \label{eq:ydot}\\
\dot{Z} &=& \Phi \mathrm{p}_z \label{eq:zdot}\,,
\end{eqnarray}
where $\Phi=v_s/U$ is the swimming number, while equation~(\ref{eq:2}) becomes
\begin{eqnarray}
\dot{\mathrm{p}}_x &=& -\frac{1}{2\Psi} \mathrm{p}_x \mathrm{p}_z 
-\frac{1}{2} \sin Z\, \mathrm{p}_z 
\label{eq:pxdot}\\
\dot{\mathrm{p}}_y &=& -\frac{1}{2\Psi} \mathrm{p}_y \mathrm{p}_z 
\label{eq:pydot}\\
\dot{\mathrm{p}}_z &=& \frac{1}{2\Psi} (1-\mathrm{p}^2_z) 
+\frac{1}{2} \sin Z\, \mathrm{p}_x \,.
\label{eq:pzdot}
\end{eqnarray}
where  $\Psi=B U/L$ is the stability number.
The box size in dimensionless units is $L_B/L=2\pi$.

The coordinates $X$ and $Y$ do not enter the dynamics of the other 
variables,
thus we can ignore them and limit our analysis to the four dimensional
dynamical system given by Eq.~(\ref{eq:zdot}) for the vertical
position and Eqs.~(\ref{eq:pxdot})-(\ref{eq:pzdot}) for the swimming
orientation. The condition $|{\bf p}|=1$ implies that the dynamics is
three-dimensional.

It is easily seen from Eqs.~(\ref{eq:pxdot}-\ref{eq:pzdot}) that when
$\Psi\leq 1$ the gravitational bias dominates allowing cells to swim
upwards through the vertical shear. Conversely, for $\Psi>1$ vorticity
becomes important inducing tumbling motion, which in turns gives rise
to gyrotactic trapping.\cite{Durham2009} In fact, seeking for an
equilibrium swimming direction in Eqs.~(\ref{eq:pxdot}-\ref{eq:pzdot})
when $\Psi>1$, a refined analysis\cite{Pedley1987} shows that, thanks
to the absence of vorticity along the direction of gravity (as for the
Kolmogorov flow here considered), the only possible equilibrium
solution is one with $\mathrm{\bf p}$ lying on the plane perpendicular
to gravity and is non-linearly unstable. Therefore, if $\Psi>1$
vortical motion overcomes gravitational bias and cells perform
tumbling motion for some height $Z$ where they remain trapped.  While
this is generic for steady shear flows with vorticity perpendicular to
gravity, for the specific case of the Kolmogorov flow it is possible
to characterize the dynamical behavior in great details. Indeed we can
notice that, besides $|{\bf p}|$,
Eqs.~(\ref{eq:xdot})-(\ref{eq:pzdot}) admit two additional conserved
quantities implying that the system is integrable.

We start our analysis by observing that, when the gravitational bias
dominates the orientation dynamics (i.e. $\Psi<1$), we expect an
average upward swimming speed $\langle \dot{Z} \rangle = \Phi \langle
\mathrm{p}_z \rangle >0$.  Since Eq.~(\ref{eq:pydot}) is formally
solved by $\mathrm{p}_y(t)=\mathrm{p}_y(0)\exp[-\int_0^t
  \mathrm{p}_z(s) ds/(2\Psi)]$, at long times we can write
$\mathrm{p}_y(t)=\mathrm{p}_y(0)\exp[-t \langle \mathrm{p}_z
  \rangle/(2\Psi)]$, meaning that asymptotically $\mathrm{p}_y \to 0$
and the swimming orientation evolves on the
$(\mathrm{p}_x,\mathrm{p}_z)$ plane. Actually we  can say more:
dividing (\ref{eq:pydot}) by (\ref{eq:zdot}) yields
$d\mathrm{p}_y/dZ=-\mathrm{p}_y/(2\Psi\Phi)$, which implies that
\begin{equation}
\mathcal{C}({\bf p},Z)= \mathrm{p}_y e^{Z/(2\Phi\Psi)} \label{eq:C1}
\end{equation}
is invariant under the dynamics (\ref{eq:zdot}-\ref{eq:pzdot}).
Furthermore, dividing (\ref{eq:pxdot}) by (\ref{eq:zdot}) and solving
the resulting ordinary differential equation for
$\mathrm{p}_x=\mathrm{p}_x(Z)$, one easily finds that
\begin{equation}
\mathcal{H}({\bf p},Z)=\Phi e^{\frac{Z}{2\Phi\Psi}} \left[\mathrm{p}_x-\frac{\Psi (2\Phi\Psi\cos Z- \sin Z)}{1+4\Phi^2\Psi^2} \right]\,, \label{eq:C2}
\end{equation}
is also conserved by the dynamics.  In
Refs.~\onlinecite{Stark_PRL2012,Stark_EPJE2013} similar considerations were
used for studying prolate cells, such as bacteria, swimming in a
Poiseuille flow.

The conservation of $\mathcal{C}$ implies that if $Z$ increases
$\mathrm{p}_y$ has to compensate decreasing exponentially, as
discussed above.  As a consequence, we can neglect, at this stage,
swimming in the $y$ direction (i.e. we set $\mathrm{p}_y=0$) and limit
our analysis to the two-dimensional system
\begin{eqnarray}
\dot{\theta} &=& \frac{1}{2\Psi} \cos\theta +\frac{1}{2} \sin Z
\label{eq:2thdot}\\
\dot{Z} &=& \Phi \sin\theta \,,
\label{eq:2zdot}
\end{eqnarray}
where we have introduced polar coordinates for the swimming
orientation, $(\mathrm{p}_x,\mathrm{p}_z)=(\cos\theta,\sin\theta)$. 

Since equations~(\ref{eq:2thdot}-\ref{eq:2zdot}) are periodic, we can
consider the evolution on the torus $(\theta,z)\in
[-\pi,\pi]\times[0:2 \pi]$.  In the following we will use $Z$ to denote
the vertical coordinate and $z$ to indicate its restriction to the
torus (i.e. $z= Z \mod 2 \pi$). We observe that the quantity
(\ref{eq:C2}) is not periodic in $Z$: when $Z \to Z\pm 2 \pi\,n$,
we have $\mathcal{H}(\theta,z) \to \mathcal{H}(\theta,z)
e^{\pm \pi n/(\Phi\Psi)}$, i.e. $\mathcal{H}$ is multiplied by a constant.

The system (\ref{eq:2thdot}-\ref{eq:2zdot}) can be
rewritten as
\begin{equation}
\begin{array}{l}
\dot{\theta}= \phantom{-}G(\theta,Z) \partial_Z \mathcal{H}\\
\dot{Z}= -G(\theta,Z) \partial_\theta \mathcal{H}
\end{array}
\label{eq:timechange}
\end{equation}
with $G=\exp[-Z/(2\Phi\Psi)]$ being the inverse integrating factor.\cite{IIF-1,IIF-2} Therefore the time change $t \to tG^{-1}$ makes
(\ref{eq:timechange}) a Hamiltonian system having exactly the same
trajectories of the original system (\ref{eq:2thdot}-\ref{eq:2zdot})
but these are traveled with different speeds. As a consequence,
while the Lebesgue measure is invariant (from Liouville theorem) for
the Hamiltonian system, this is not the case for
Eqs.~(\ref{eq:2thdot}-\ref{eq:2zdot}), which explains why one can
observe accumulations of swimmers (see below).

\begin{figure}[t!]
\centering
\includegraphics[width=.5\textwidth]{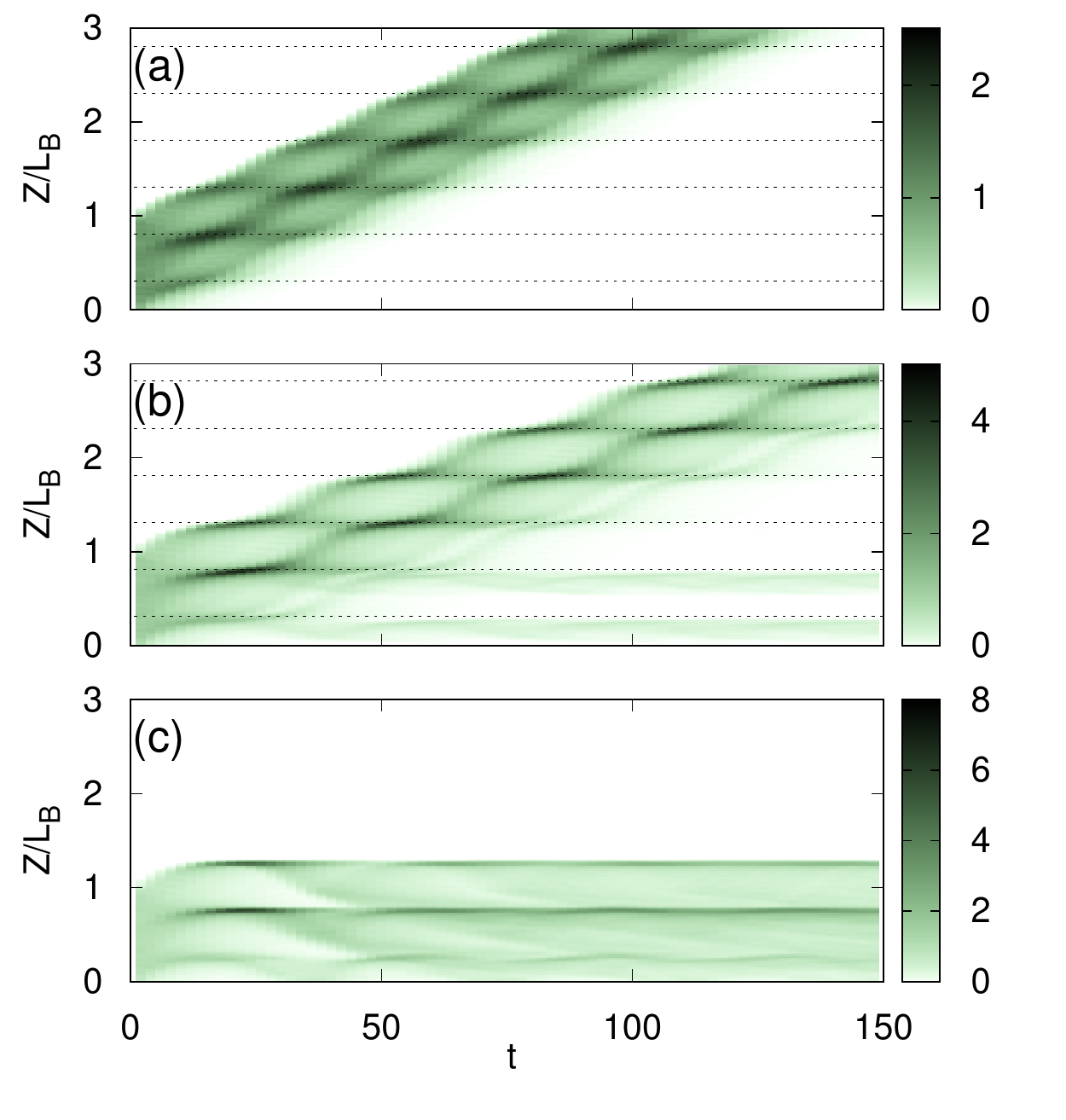}
\caption{(color online) Evolution of the vertical density of cells,
  $\rho(Z,t)$, in the 2d laminar Kolmogorov flow for $\Phi=0.2$ with
  (a) $\Psi=0.9<\Psi_c$, (b) $1<\Psi=1.06<\Psi_c$ and
  $\Psi=1.12>\Psi_c$, where $\Psi_c$ is given by
  (\protect\ref{eq:psic}).  Dotted horizontal lines in (a) and (b)
  mark vertical velocity minima (see text for a discussion).  The
  density has been obtained coarse-graining the vertical position of
  $N=10^4$ cells initialized uniformly in $(\theta,Z)\in
  [-\pi,\pi]\times[0:2\pi]$ and evolved by integrating
  Eqs.~(\ref{eq:2thdot}) and (\ref{eq:2zdot}) with a 4$^{th}$-order
  Runge-Kutta scheme.  
\label{fig:denslam}}
\end{figure}

Let us now inspect Eqs.~(\ref{eq:2thdot}-\ref{eq:2zdot}) more closely.
For $\Psi<1$ the dynamics does not admit fixed points, and $Z$ 
grows in time, as discussed above.
The conservation of $\mathcal{H}$, and in particular the
exponential dependence on $Z$, implies that when cells migrate upwards
the term in square brackets in (\ref{eq:C2}) must decrease
exponentially with the vertical position to keep
$\mathcal{H}=const$. Therefore, for large $Z$, the swimming direction
will be given by
\begin{equation}
\cos\theta=\mathrm{p}_x=\frac{\Psi (2\Phi\Psi\cos Z- \sin Z)}{1+4\Phi^2\Psi^2}\,.
\label{eq:pxasym}
\end{equation}
Remarkably, because $\mathrm{p}_z=\sqrt{1-\mathrm{p}_x^2}$ depends on
$Z$, the vertical velocity will change with height and cells will
accumulate where it is minimal.  A straightforward computation shows
that the minima of $\mathrm{p}_z$ occur at
$Z=n\,\pi-\arctan[1/(2\Phi\Psi)]$, for any integer $n$.  Around these
positions one expects to observe ephemeral layers (also for $\Psi<1$)
of high density of cells.  The transient accumulations last longer for
smaller values of the swimming number $\Phi$. The above picture is
confirmed in Fig.~\ref{fig:denslam}a showing the time evolution of the
vertical probability density distribution (PDF), $\rho(Z,t)$,
resulting from an initially uniform distribution in $Z \in [0:2\pi]$
for $\Psi<1$.

The above scenario applies whenever Eq.~(\ref{eq:pxasym}) holds
globally, i.e. for any values of $Z$. As $|\mathrm{p}_x| \le 1$, it is
easy to see that for swimmers with $\Phi<\Phi_c=1/2$,
Eq.~(\ref{eq:pxasym}) can be satisfied only if $\Psi\leq \Psi_c$ with
\begin{equation}
\Psi_c= (1-4\Phi^2)^{-1/2}\,.\label{eq:psic}
\end{equation}
For fast enough swimmers, $\Phi\geq \Phi_c$, Eq.~(\ref{eq:pxasym})
holds for any value of the stability number $\Psi$.
As $\Psi_c \ge 1$, we must distinguish two cases.

In the first (Fig.~\ref{fig:denslam}b), $1<\Psi<\Psi_c$, a
fraction of cells migrate upwards asymptotically setting their motion
on the orbit (\ref{eq:pxasym}), as in Fig.~\ref{fig:denslam}a for
cells with $\Psi<1$. Also in this case maxima of cell density
correspond to minima of $\mathrm{p}_z$. However, unlike the case
$\Psi<1$, we now observe a non negligible fraction of cells (depending
on the initial conditions) which do not migrate upwards and accumulate
in thin layers, now not in correspondence of the minima of
$\mathrm{p}_z$ (Fig.~\ref{fig:denslam}b).

In the second, $\Psi>\Psi_c$, we observe that all swimmers become
trapped and generate an inhomogeneous vertical density profile which
soon becomes stationary in time and organized in thin layers
(Fig.~\ref{fig:denslam}c).

\begin{figure}[t!]
\centering
\includegraphics[width=.23\textwidth]{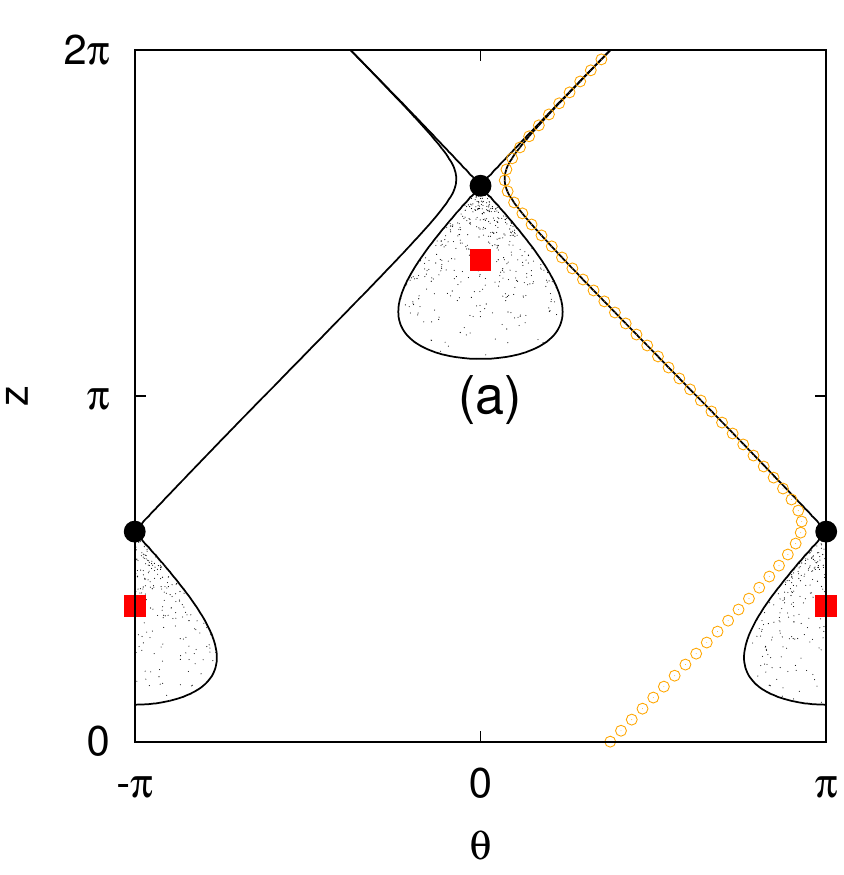} \includegraphics[width=.23\textwidth]{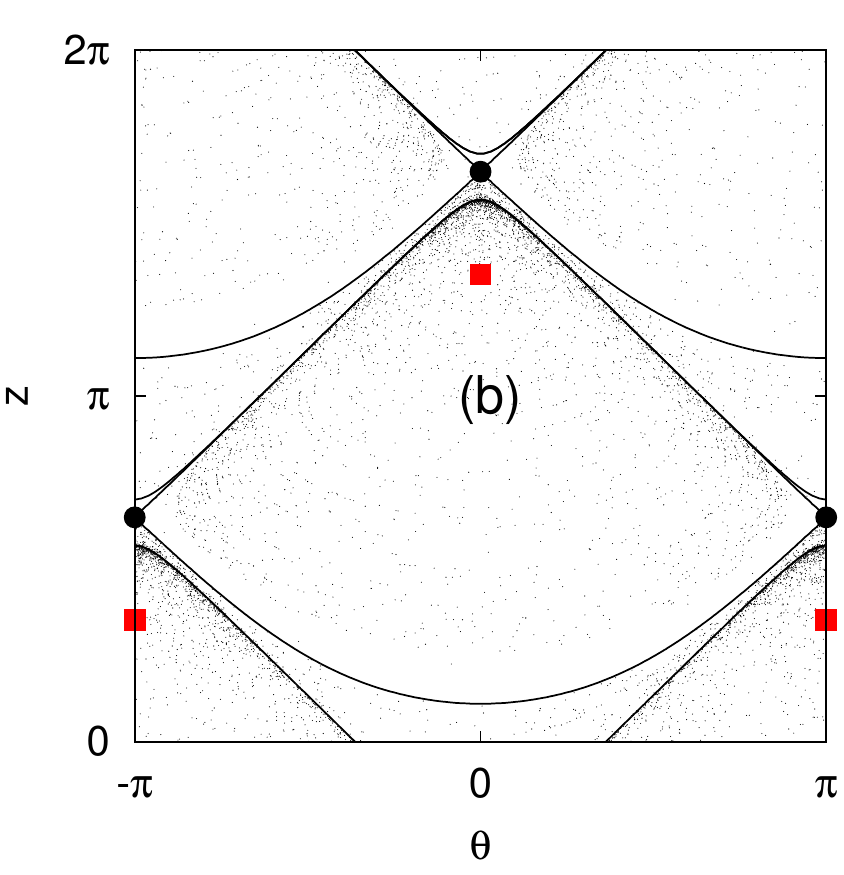}
\caption{Cells positions in the laminar 2D Kolmogorov flow at a long
  time on the $(\theta,z)$ torus for (a) $\Psi=1.06$ and (b)
  $\Psi=1.12$ with $\Phi=0.2$, corresponding to
  Fig.~\ref{fig:denslam}b and c, respectively.  Black circles (red
  squares) mark the hyperbolic (elliptic) fixed points
  (\protect\ref{eq:fixedpoints}). Black curves denote the separatrices
  emerging from the hyperbolic fixed points, which are obtained from
  the isolines of $\mathcal{H}$ computed at $H1$ and $H2$.  At
  $\Psi<\Psi_c$ (a) small black dots corresponds to trapped orbits and
  empty circles to the orbit (\protect\ref{eq:pxasym}) which
  asymptotically characterized the vertically migrating cells and is
  very close to the separatrix.  For $\Psi>\Psi_c$ (b) all orbits are
  trapped.
\label{fig:sep}}
\end{figure}

We can understand the different behaviors observed in
Fig.~\ref{fig:denslam}b and Fig.~\ref{fig:denslam}c by inspecting the
phase-space qualitative dynamics for $1<\Psi<\Psi_c$ and
$\Psi>\Psi_c$.  First, we observe that, for $\Psi>1$,
Eqs.~(\ref{eq:2thdot}-\ref{eq:2zdot}) admit the following fixed points
(written in the reference torus)
\begin{equation}
(\theta^*,z^*)=
\left\{
\begin{array}{ll}
H1  & (0,2\pi-\arcsin(\Psi^{-1}))\\
H2 & (\pi,\pi-\arcsin(\Psi^{-1})) \\
E1 & (0,\pi+\arcsin(\Psi^{-1})) \\
E2 & (\pi,\arcsin(\Psi^{-1}))\,,
\end{array}
\right. \label{eq:fixedpoints}
\end{equation}
where $H$'s and $E$'s are hyperbolic and elliptic fixed points (see
Fig.~\ref{fig:sep}) with eigenvalues $\lambda=\pm
(\Phi\sqrt{1-\Psi^{-1}}/2)^{1/2}$ and $\pm i
(\Phi\sqrt{1-\Psi^{-1}}/2)^{1/2}$, respectively.

\begin{figure}[b!]
\centering \includegraphics[width=0.45\textwidth]{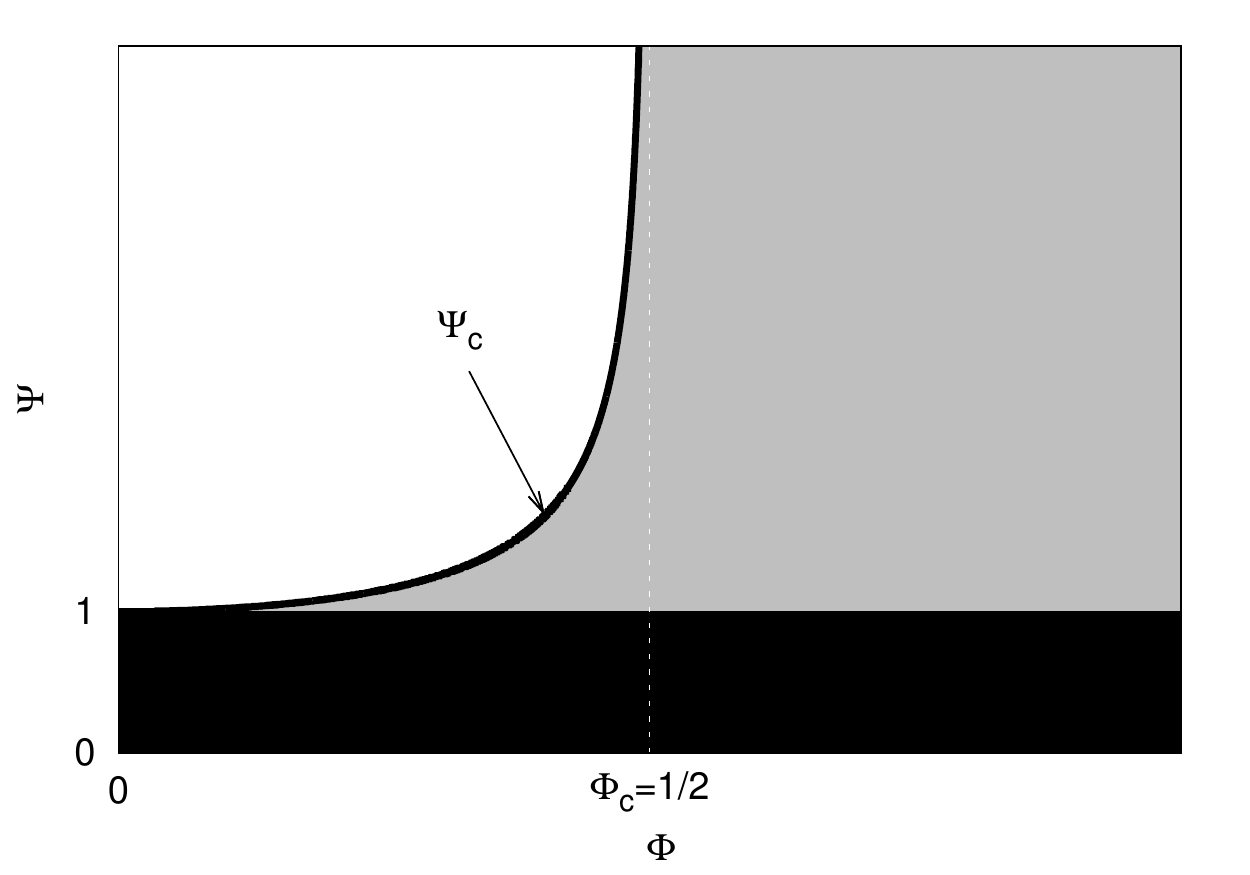}
\caption{Behavior of swimming cell in parameter space for the laminar
  Kolmogorov flow.  The white region corresponds to vertically trapped
  orbits ($\Psi>\Psi_c$), the grey one to partially trapped
  trajectories ($1<\Psi< \Psi_c$, with coexistence of trapped and
  vertically migrating cells) and, finally, the black to vertically
  migrating cells $(\Psi<1)$, whose swimming direction depends on the
  vertical position as predicted by Eq.~(\protect\ref{eq:pxasym}).
\label{fig:regions}
}
\end{figure}

Below and above the critical value $\Psi_c$ the form of the
separatrices, i.e. the orbits emerging from the hyperbolic fixed
points, changes qualitatively.  For $1<\Psi<\Psi_c$ the separatrices
roll up vertically around the torus with a slip-knot from the
hyperbolic point containing the elliptic one (see
Fig.~\ref{fig:sep}a).  Orbits initially within the slip-knot remain
trapped there, while those starting outside the slip-knot migrate
vertically, asymptotically following the orbit
(\ref{eq:pxasym}). Conversely, when $\Psi>\Psi_c$, the separatrices
roll up around the torus in the $\theta$ direction
(Fig.~\ref{fig:sep}b) acting as barriers to vertical transport as
typically happens in Hamiltonian
systems.\cite{lichtenberg1992}  Hence, whenever $\Psi>\Psi_c$
trajectories remain bounded in the vertical direction for all initial
conditions. It is noted that for $\Psi=\Psi_c$ the orbit
(\ref{eq:pxasym}) becomes the separatrix and passes through all the
hyperbolic points.  Figure~\ref{fig:regions} summarizes the possible
behaviors in parameter space $(\Phi,\Psi)$.

\begin{figure}[t!]
\centering
\includegraphics[width=.45\textwidth]{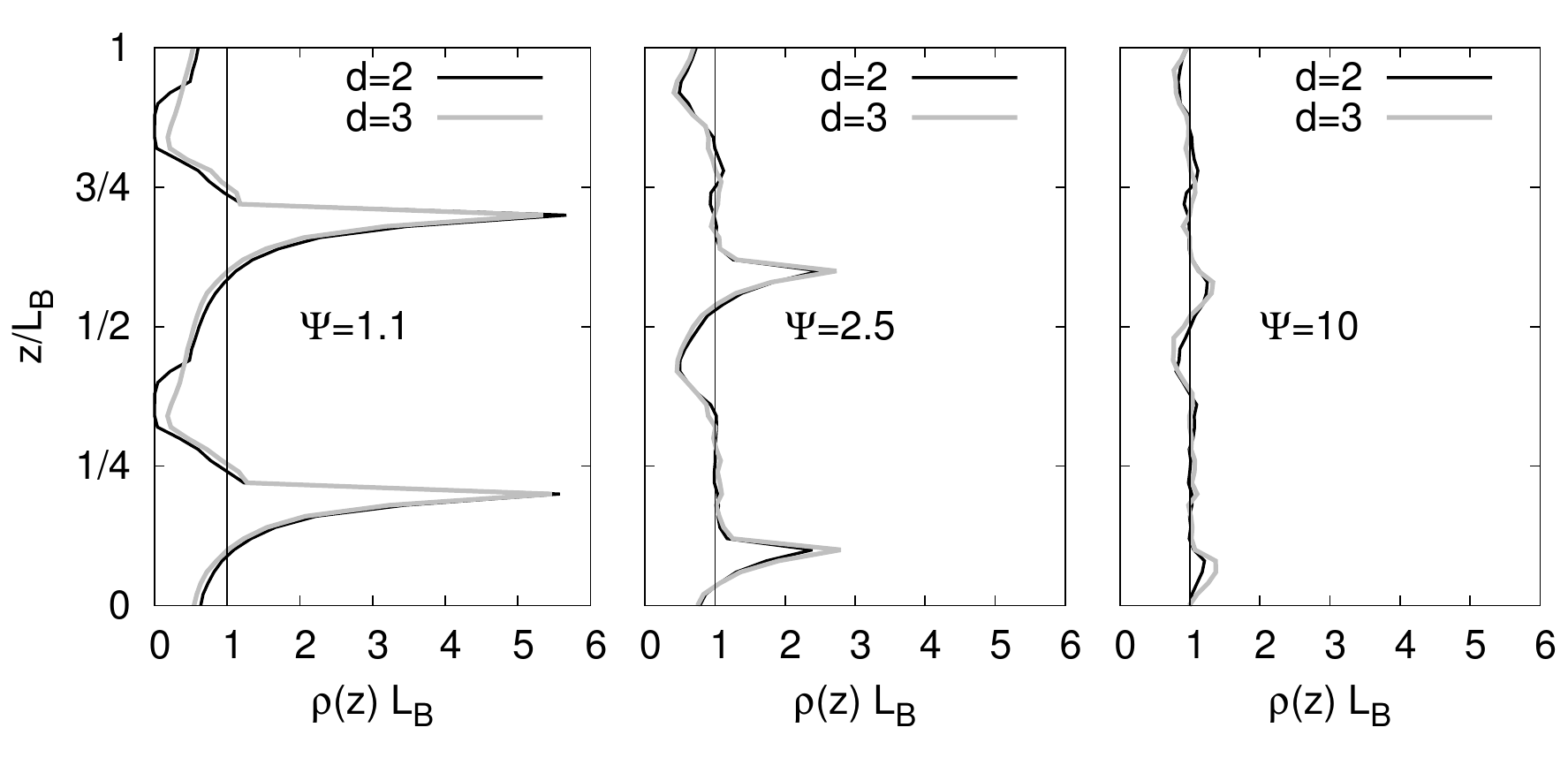}
\caption{Number density of algae corresponding to the two layers
  restricted to the domain $z \in[0:L_B]$, for both 2D and 3D for
  $\Phi=0.05$ and three values of $\Psi$ as labeled. Density differs
  for minor details in the two cases. Notice that layers tend to
  disappear for large $\Psi$.
\label{fig:rhoz-lam}}
\end{figure}

It should be noted that this layered structure is essentially due to
the fact that: $(i)$ the velocity on the orbit depends on the vertical
position, and the trajectory spends more time where $G$ is smaller (i.e. $Z$
large) as from Eq.~(\ref{eq:timechange}); $(ii)$ the fact that the
separatrices confine the motion. It is easy to understand that $(i)$
and $(ii)$ imply that concentration will be large around the highest
allowed vertical value which, for $1<\Psi<\Psi_c$
(Fig.~\ref{fig:sep}a), coincides with the hyperbolic points and, for
$\Psi>\Psi_c$ (Fig.~\ref{fig:sep}b), is in between the hyperbolic and
elliptic points. Cell accumulation will thus increase going upwards to
the top of the separatrices and then will abruptly fall down, as
revealed by the vertical asymmetry in density profiles shown in
Fig.~\ref{fig:rhoz-lam}, see also Ref.~\onlinecite{Durham2012} for a
discussion on such asymmetries.

It should be noted that at increasing $\Psi$, the accumulation in
layers tends to disappear (see also Fig.~\ref{fig:rhoz-lam}).  Indeed
in the limit $\Psi\to \infty$ the dynamics
(\ref{eq:2thdot}-\ref{eq:2zdot}) becomes Hamiltonian with
$\mathcal{H}=\Phi\cos\theta-(1/2)\cos(Z)$, corresponding to the well
known Harper Hamiltonian, originally introduced to describe crystal
electrons in the presence of a magnetic field.\cite{Harper1955}
Consequently, swimming cells cannot display accumulation as implied by
Liouville theorem.  Nonetheless, provided $\Phi<\Phi_c$, also in this
limit we have that the separatrices act as barriers to vertical
migration.

We conclude this section briefly commenting the three dimensional
case. As discussed above, thanks to the invariance of (\ref{eq:C1})
upward migrating cells (as in the black and grey region of
Fig.~\ref{fig:regions}) follow trajectories which a long times
coincide with the two dimensional case (as $\mathrm{p}_y\to 0$).  In
principle, the dynamics of non-vertically-migrating cells can be fully
characterized, e.g., studying the conserved quantities.\cite{Stark_PRL2012} On a qualitative level, the orientation vector
will move on the intersection between the surface determined by
(\ref{eq:C1}) and (\ref{eq:C2}) and the sphere $|{\bf p}|=1$. The
conservation of (\ref{eq:C1}) implies that the extreme values reached
by $\mathrm{p}_y$ are linked to the extremes of $Z$. The same conservation law
also shows that $\mathrm{p}_y$ is bounded away from $0$ for any finite $Z$ and
any initial condition with $\mathrm{p}_y\neq 0$. As a consequence, all bounded
trajectories will have a finite average drift along $y$, according to
the initial sign of $\mathrm{p}_y$. However, since $y$ is a slaved variable,
the above depicted scenario, including the behavior in parameter space
(Fig.~\ref{fig:regions}) is unmodified going from two to three
dimensions. Actually, as shown in Fig.~\ref{fig:rhoz-lam} the
(vertical) density profiles are quantitatively very close in two and
three dimensions. 

\section{Effects of rotational diffusion\label{sec:4}}

In this section we focus on the effects of
  stochasticity on the dynamics of the swimming orientation. Even in
a still fluid, indeed, swimming trajectories are not
straight lines and usually display a certain degree of randomness
because of thermal fluctuations and/or of the swimming
process. Thermal fluctuations are important for very small ($\sim 1\mu
m$) microorganisms, such as e.g. bacteria, and can be modeled in terms
of rotational Brownian motion (RBM),\cite{Berg2004} namely as a
diffusion of the swimming direction on the unit sphere.  Gyrotactic
microalgae are typically too large to be affected by thermal
fluctuations. However, in theoretical
approaches\cite{Pedley1992,Bees1998} and consistently with
experimental observations,\cite{hill97,Vladimirov2004} it is still
possible to use RBM to model the random fluctuations of the swimming
direction due, e.g., to small variations in the cell shape, waving or
imperfections in the flagella movement, and bacteria-like
run-and-tumble\cite{Polin2009} due to the desynchronization between
the flagella.

\begin{figure}[t!]
\centering
\includegraphics[width=0.5\textwidth]{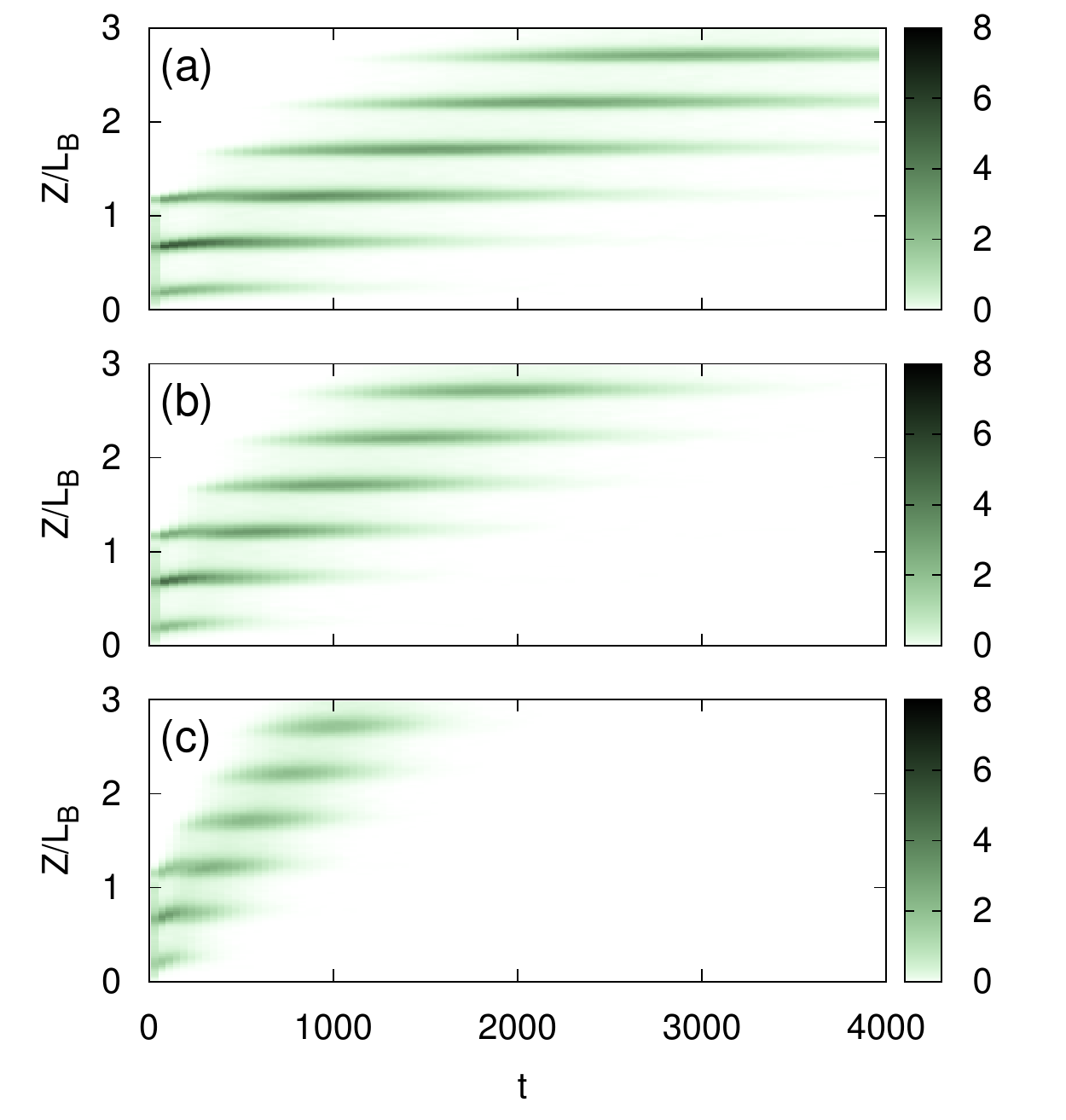}
\caption{(color online) Evolution of the vertical number density
  $\rho(Z,t)$ in the 2D laminar Kolmogorov flow with parameter as in
  Fig.~\ref{fig:denslam}c and the presence of rotational Brownian
  motion with (a) $Pe_{r}^{-1}=0.005$, (b) $0.01$, and (c) $0.04$. 
\label{fig:densrot}}
\end{figure}

For the sake of simplicity, we only consider the 2D Kolmogorov flow,
which displays, also quantitatively (see, e.g.,
Fig.~\ref{fig:rhoz-lam}), the main features of the 3D flow.  In two
dimensions, RBM corresponds to diffusion of the angle $\theta$ in
Eq.~(\ref{eq:2thdot}) with (rotational) diffusivity $D_r$.  As an
example, in \textit{Chlamydomonas
  augustae}\cite{williams2011} $D_r^{-1}\! \sim\!
  15\, s$. It is useful to introduce a non-dimensional measure of
  diffusion, namely the rotational Peclet number
  $Pe_{r}=U/(LD_r)$. Considering RBM amounts to
  adding to the r.h.s. of (\ref{eq:2thdot}) the
  stochastic term $\sqrt{2Pe_{r}^{-1}}\eta$, $\eta$ being a zero-mean
  Gaussian variable with $\langle\eta(0)\eta(t)\rangle=\delta(t)$.

RBM can cause dramatic effects when gyrotactic trapping is effective
(i.e. when $\Psi>\Psi_c$): in principle, for any $Pe_r^{-1}>0$, thanks
to random fluctuations, all swimmers potentially have a way to escape
from the ``barrier'' of the separatrices (by definition, impenetrable
in the deterministic case).  We therefore expect that a small random
component in the swimming dynamics will make layers transient, with a
finite lifetime.

This scenario is confirmed in Fig.\ref{fig:densrot}
showing the time evolution of the vertical cell-number density for
stability number corresponding to gyrotactic trapping
(with $\Psi>\Psi_c$ as in Fig.~\ref{fig:denslam}c)
when RBM is acting on the dynamics.  Similarly to the deterministic
case, an initially uniform distribution in $(\theta, Z) \in
[-\pi:\pi]\times[0:L_B]$ (quickly) evolves into layers located around
the elliptic points. However, unlike the deterministic case, after a
typical time depending on $Pe_{r}$, any layer dissolves and, thanks to
the flow periodicity, gives birth to a new layer
at distance $L_B/2$ upward. The ``length'' of the traces in
Fig.~\ref{fig:densrot} essentially corresponds to the lifetime of a
layer.  Clearly, in the case of a non-periodic set-up only a single
layer would form, persist for some time and then dissolve unless a
continuous in-flow of algae is provided from below.
We notice that the lowest layer lasts for about half the duration of 
the other layers: this is due to the fact that it has no layers below feeding
it. 
The figure shows that the value of $Pe_{r}$ influences both
the lifetime and the focusing of the layers. We shall be more
quantitative on this aspect in the following.

\begin{figure}[t!]
\centering
\includegraphics[width=.5\textwidth]{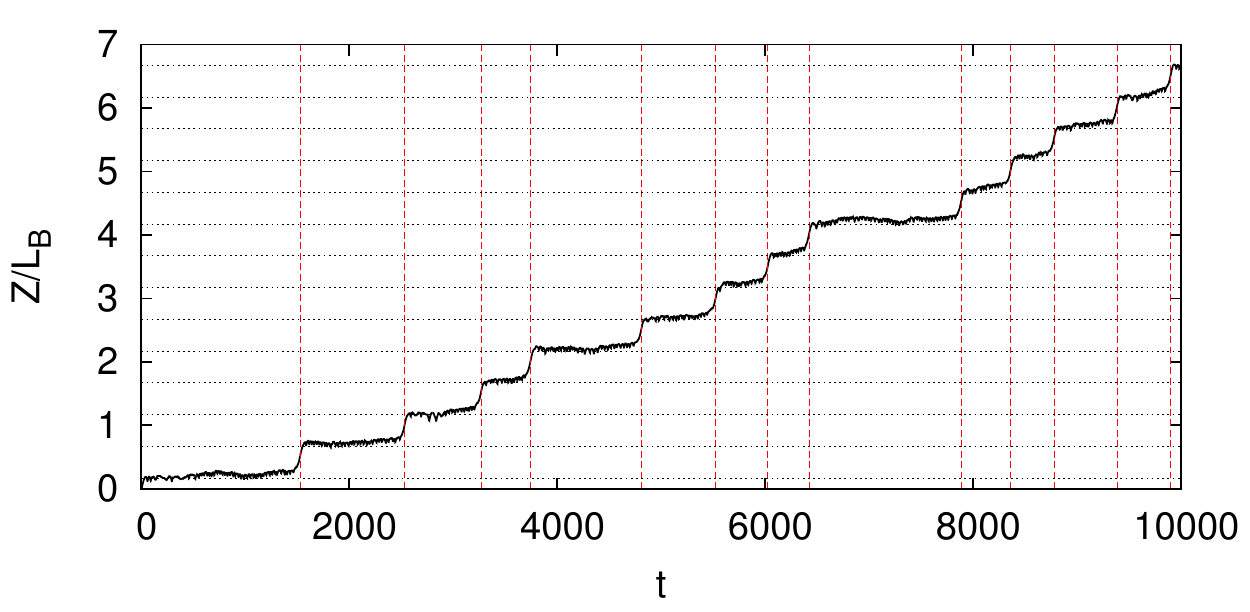}
\caption{(Color online) Typical cell trajectory for $\Phi=0.05$,
  $\Psi=1.12$ and $Pr_r^{-1}=0.005$.  We can identify two regimes:
  temporary trapping in a layer (around the horizontal dashed lines
  demarking the elliptic points) and short jumps between layers.
  Vertical (red) lines mark the transition between adjacent layers,
  numerically identified as the times at which the vertical position
  $Z$ increases of $L_B/2$.  \label{fig:trajrot}}
\end{figure}

\begin{figure}[t!]
\centering
\includegraphics[width=0.5\textwidth]{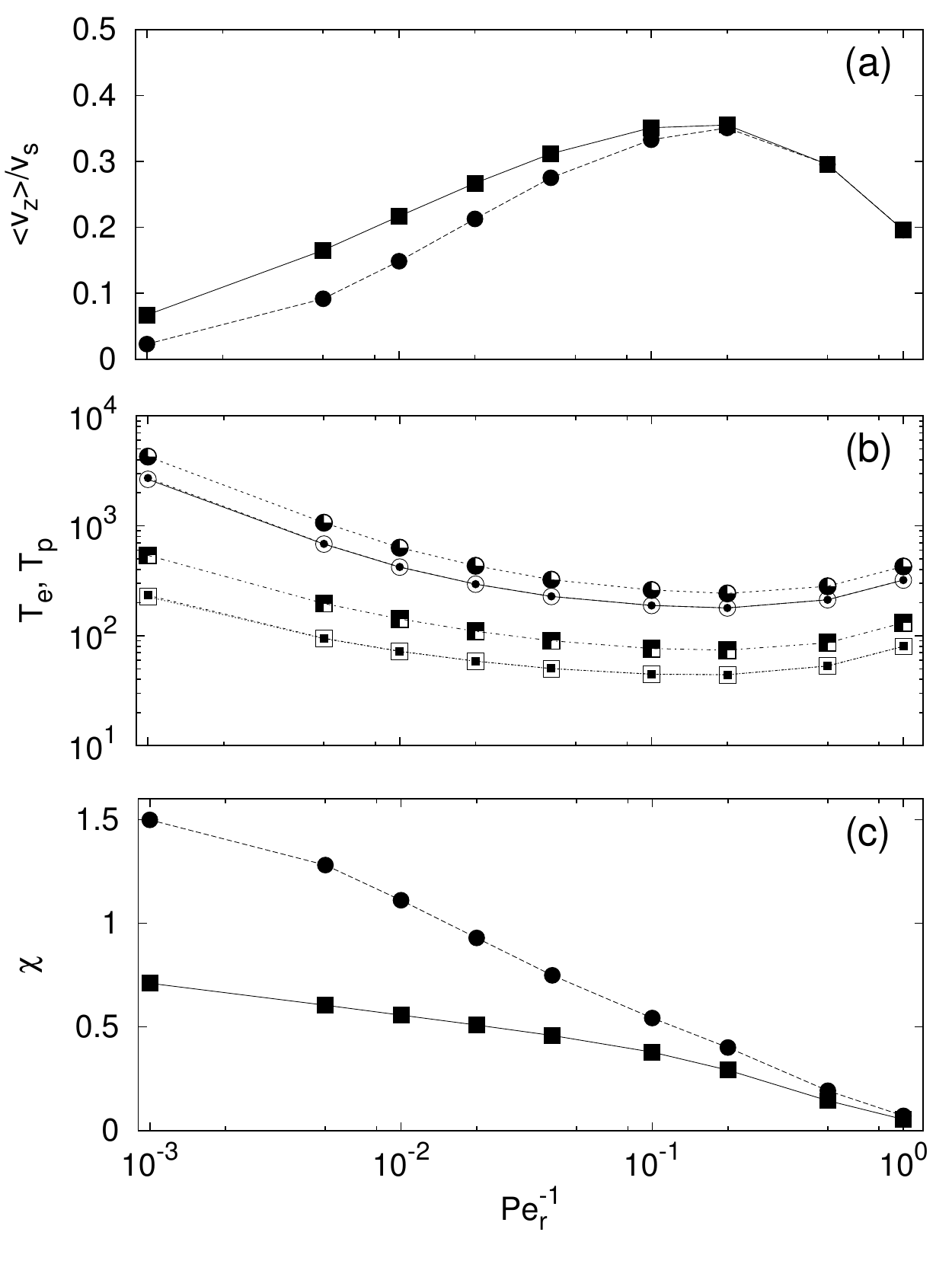}
\caption{Quantitative characterization of single cell and layer
  properties as a function of the rotational Peclet number $Pe_r$, for
  two values of the swimming parameter $\Phi=0.05$ (circles) and
  $\Phi=0.2$ (squares) at $\Psi=1.12>\Psi_c$ in the 2D laminar
  Kolmogorov flow.  (a) Time and population average vertical velocity
  $\langle v_z\rangle$ normalized to the swimming speed $v_s$.  (b)
  Average trapping time $T_e$ (open symbols) for single trajectories
  and layer persistence time $T_p$ (semi-filled symbols). Notice that
  the exit time $T_e$ coincides with $L_B/(2\langle v_z\rangle)$ (small
  filled symbols inside the empty ones).  
  (c) Inhomogeneity index $\chi$ defined in (\ref{eq:chi}).}
\label{fig:medie}
\end{figure}

To better understand the process it is useful to inspect the behavior
of a typical single swimmer trajectory (Fig.~\ref{fig:trajrot}). We
can clearly identify trapped states interrupted by rapid upward
migrations.  As from Fig.~\ref{fig:sep}b, trapping is spatially
localized at $Z$ values (mainly) in between adjacent elliptic and
hyperbolic points, where a swimmer can spend a long time before RBM
allows it to escape the separatrix. Out of the trapping region, shear
vorticity is low and thus the cell can locally migrate vertically (as
globally done when $\Psi<1$), till it enters a new trapping region
shifted of $L_B/2$ above due to the flow periodicity. Then the process
starts again. The (stochastic) switch between these two states of
motion induces an average vertical drift, $\langle v_z \rangle>0$
(Fig.~\ref{fig:medie}a). Clearly, in both the deterministic
($Pe^{-1}_r\to 0$) and RBM dominated (large $Pe_r^{-1}$) we should
expect a zero average drift: in the former case because of gyrotactic
trapping, in the latter due to fast
decorrelation of the swimming orientation due to RBM (this is expected
when $\Psi/Pe_r=BD_r$ becomes very large). 
Consequently, we expect to have an intermediate value of the rotational 
diffusivity for which the
vertical drift is maximal, as confirmed in Fig.~\ref{fig:medie}a.

The qualitative features of the trajectory shown in
Fig.~\ref{fig:trajrot} suggest to look at the statistics of trapping
time $T$. Thanks to the periodicity and the fact that out of the layer
vertical migration is fast (Fig.~\ref{fig:trajrot}), we can define it
as the time $T$ it takes for a swimmer to swim upwards the distance
between two consecutive layers (i.e. $L_B/2$ in our model flow). The
average exit (or trapping) time $T_e=\langle T \rangle$ shown in
Fig.~\ref{fig:medie}b coincides with $L_B/(2\langle v_z
\rangle)$. While this is obvious as the two statistics are
mathematically equivalent, it can be useful when coping with finite
observation times. In fact, exit-time statistics can be strongly
biased when the total observation time is not large enough (see
Sect.~\ref{sec:5}).
  
A quantitative comparison between the average exit time $T_e$ of the
single trajectory and Fig.~\ref{fig:densrot} suggests however that
$T_e$ tends to underestimate the layers time duration. A more sounding
definition of the layer persistence time, $T_p$, requires somehow to
account for the fact that there could be many swimmers trapped for
times longer than the average $T_e$. Heuristically we found that a
reasonable estimate is obtained considering that a layer dissolves
when, say, $\sim 90\%$ of the cells have escaped from it. In terms of
the exit time probability density function $p(T)$ we can thus define
$T_p$ implicitly as $\int_0^{T_p} p(T) dT \approx
0.9$. Fig.~\ref{fig:medie}b shows also $T_p$, which appears to be a
better proxy of the persistence time, for instance in the case of
Fig.~\ref{fig:densrot}b, i.e. $Pe_r^{-1}=0.01$, we have $T_p\approx
630$ which is essentially the trace length of the lowest layer,
i.e. half of the other layers.

The interpretation of $T_p$ as the layer persistence time, however,
becomes meaningless when layers are not well defined. For example,
from Fig.~\ref{fig:medie}b one can get the wrong impression that at
large $D_r$ layers last longer and longer. In reality,
by increasing $D_r$ (i.e. decreasing the coherence in
the swimming orientation) layers spread away, becoming less and
less well defined as high cell density locations (as when $\Psi$
becomes too large, see Fig.~\ref{fig:rhoz-lam}). To quantify such an
effect, we introduce a measure based on the quadratic deviation from
the uniform distribution on the domain $z\in [0:L_B]$. In other
terms we use the periodicity to restrict the vertical position in
$z\in [0:L_B]$ so that we can define the density $\rho(z,t)$. With
RBM such density reaches a statistically stationary profile and we
define the normalized root mean square deviation of the average
profile, $\rho(z)$, from that expected for a uniform distribution,
i.e. $\rho(z)=\rho_0=1/L_B$. In formulae, this \textit{inhomogeneity
index} is defined as
\begin{equation}
\chi={\sqrt{\langle\left(\rho-\rho_0\right)^2\rangle}}/{\rho_0}\,,
\label{eq:chi}
\end{equation}
the angular brackets denoting integration over $z$.  Figure
\ref{fig:medie}c shows that this measure monotonically decreases with
$D_r$, as expected. Also, faster swimmers concentrate less.

\section{Swimming in the turbulent Kolmogorov flow\label{sec:5}}

As discussed in Sect.~\ref{sec:2.2}, the steady Kolmogorov flow
becomes unstable for ${\rm Re}>\sqrt{2}$ and ultimately turbulent,
upon further increasing ${\rm Re}$. Nonetheless, thanks to the
monochromatic character of the (time-averaged) mean flow, we can
always decompose velocity and vorticity, entering
Eqs.~(\ref{eq:1})-(\ref{eq:2}), in the mean shear with superimposed
fluctuations $\bm u^\prime$ and $\bm \omega^{\prime}$ as
\begin{eqnarray}
\bm u &=&\phantom{-} U\cos(z/L) \hat{\bm x} + \bm u^\prime({\bm x},t) \label{eq:u}\\
\bm \omega &=& -\frac{U}{L}\sin(z/L)\hat{\bm y} + \bm \omega^\prime({\bm x},t) \,.\label{eq:w}
\end{eqnarray}
Such decomposition suggests that we can consider turbulent
fluctuations as a perturbation of the dynamics studied in
Sect.~\ref{sec:3}. Actually, even at relatively low $Re$, the
amplitude of turbulent fluctuations is of the same
  order as the mean flow, in particular $u^\prime_{\rm rms}/U\simeq
0.5$.\cite{Musacchio2014} In real oceans, however, fluctuations are
typically smaller than the mean flow due to different factors such as,
e.g., stratification.\cite{Thorpe2007}

In the following we will therefore consider the velocity (and
vorticity) field defined as $U\cos(z/L) \hat{\bm x} +\gamma\bm
u^\prime$, where $\bm u^\prime$ is the fluctuating component in
(\ref{eq:u}), obtained by a direct numerical simulation (DNS) of
Eq.~(\ref{eq:ns}).  In this way, the parameter $\gamma$ controls the
intensity of turbulent fluctuations so that $u_{\rm
  rms}^\prime/U\simeq 0.5\gamma$.  The ability to
control the weight of fluctuations is important to systematically
assess the role of fluctuations. Indeed, there are
indications\cite{brandt} that in the standard Kolmogorov flow
(i.e. $\gamma=1$), turbulence is so intense to
completely dissolve phytoplankton layers.  Another advantage of this
approach is that the statistical properties of the turbulent
fluctuations do not change with $\gamma$ as they would, for example,
by introducing stratification.

\begin{figure}[b!]
\centering
\includegraphics[width=.5\textwidth]{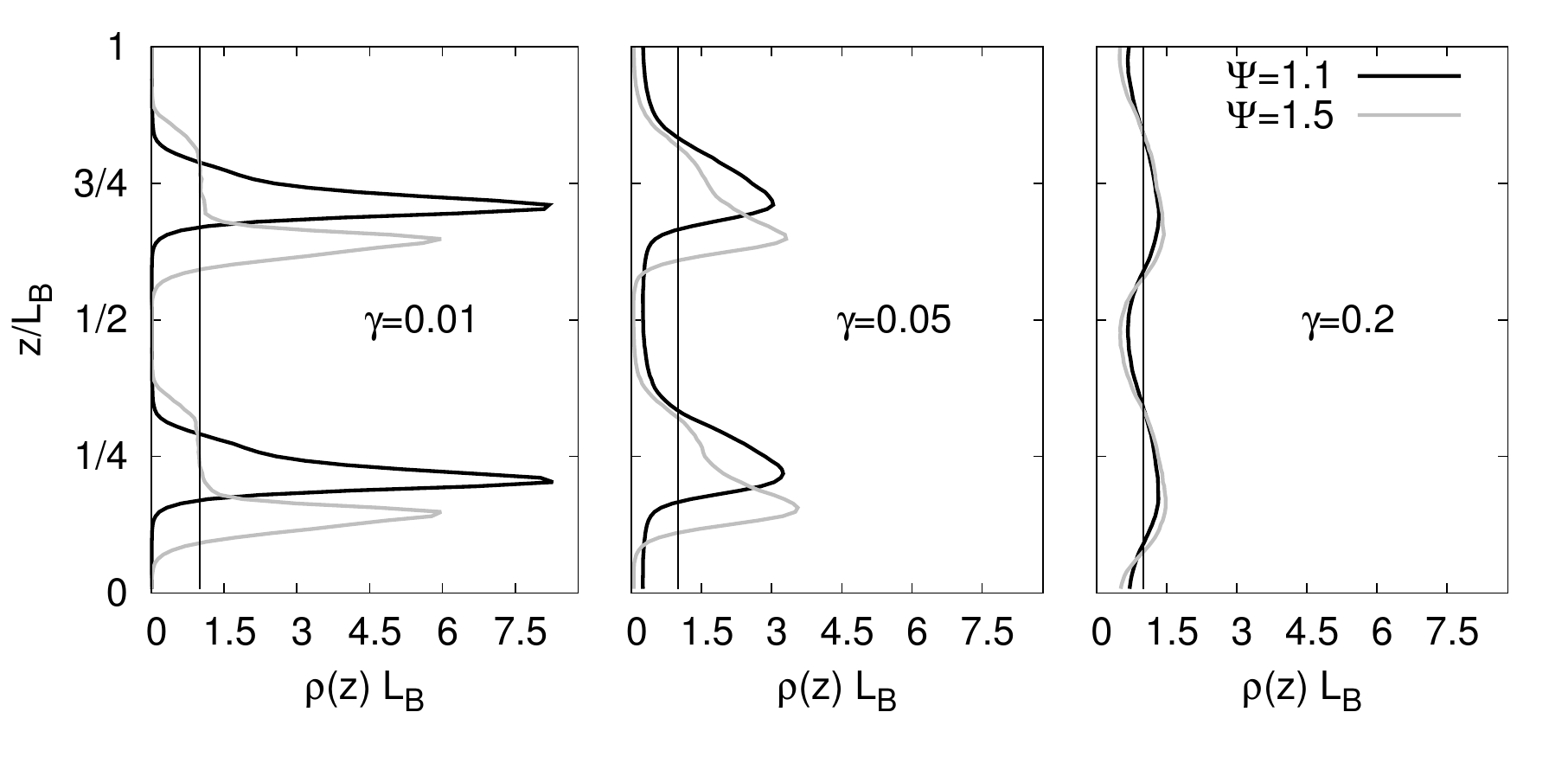}
\caption{Vertical profile of swimmer number density $\rho(z)$
for two values of the stability parameter $\Psi=1.1$ (black lines)
and $\Psi=1.5$ (grey lines) at different turbulent intensities 
$\gamma=0.01$, $\gamma=0.05$ and $\gamma=0.2$ (from left to right).
\label{fig:rhoz-turb}}
\end{figure}

Navier-Stokes equations~(\ref{eq:ns}) are integrated by means of 
a standard, fully
parallel pseudo-spectral code\cite{Musacchio2014} on a cubic domain
of size $L_B$ discretized by $128^3$ grid points, 
with periodic boundary conditions
in all directions. Lagrangian dynamics (\ref{eq:1}-\ref{eq:2})
(suitable modified with the $\gamma$-factor as discussed above) of up
to $10^5$ swimmers for each set of parameters ($\Phi,\Psi,\gamma$) was
performed using linear interpolation of the velocity and vorticity
fields, as in Refs.~\onlinecite{Durham2013,DeLillo2014}.  A $2^{nd}$-order
Runge-Kutta scheme was used for time advancement of both Eulerian and
Lagrangian dynamics.  Particle positions were re-boxed within the periodic
domain in the $x$ and $y$ directions, while the absolute displacement was
tracked along $z$.  The Reynolds number in our simulation is $Re=158$, the
smallest Kolmogorov scale $\eta$ is well resolved as $k_{\rm max}\eta\approx
1.8$.

\begin{figure}[t!]
\centering
\includegraphics[width=0.5\textwidth]{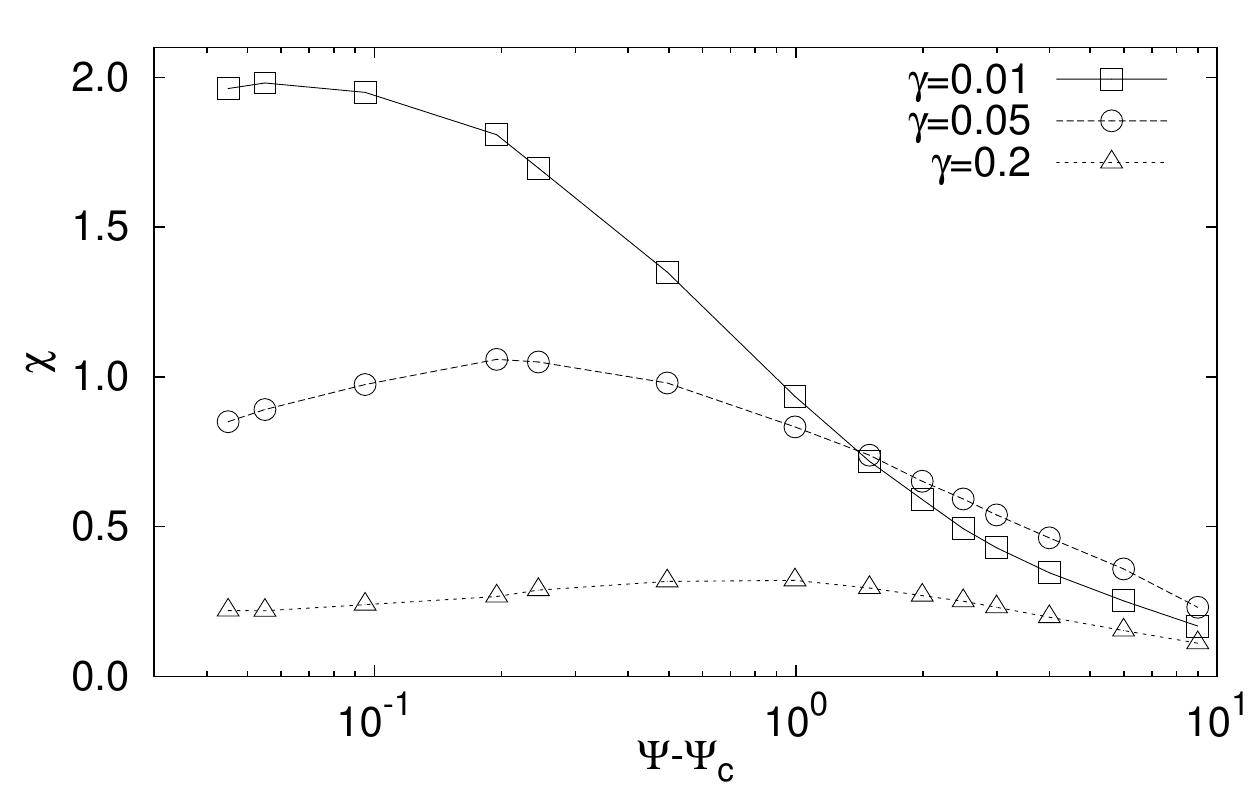}
\caption{Inhomogeneity index $\chi$ as a function of $\Psi$ for
  various $\gamma$.  Data symbols are shown only for values of the
  stability number such that gyrotactic trapping is effective in the
  laminar case, i.e. $\Psi>\Psi_c$.
\label{fig:chi-turb}}
\end{figure}

Turbulent fluctuations, when strong enough, inhibit gyrotactic
trapping. This is confirmed in Fig.~\ref{fig:rhoz-turb}, showing the
swimmer vertical-density profiles for $\Psi=1.1$ and $\Psi=1.5$ at
varying the intensity of turbulent fluctuations $\gamma$. Upon
increasing $\gamma$ vertical heterogeneity weakens. This effect is
quantified by the inhomogeneity index $\chi$ Eq.~(\ref{eq:chi}), shown
in Fig.~\ref{fig:chi-turb} as a function of the stability parameter
$\Psi$ for three values of the turbulent intensity $\gamma$.  As in
the previous Section, we focus here on the case $\Psi>\Psi_c$, when
gyrotactic trapping is effective in the laminar regime.
Interestingly, at increasing $\gamma$ turbulent fluctuations not only
smooth the inhomogeneities (decreasing the value of $\chi$) but also
induce a non-monotonic dependence on $\Psi$, with maximal
inhomogeneity obtained for a value of the stability parameter, $\Psi
\gtrsim \Psi_c$, weakly depending on $\gamma$.

\begin{figure}[t!]
\centering
\includegraphics[width=0.5\textwidth]{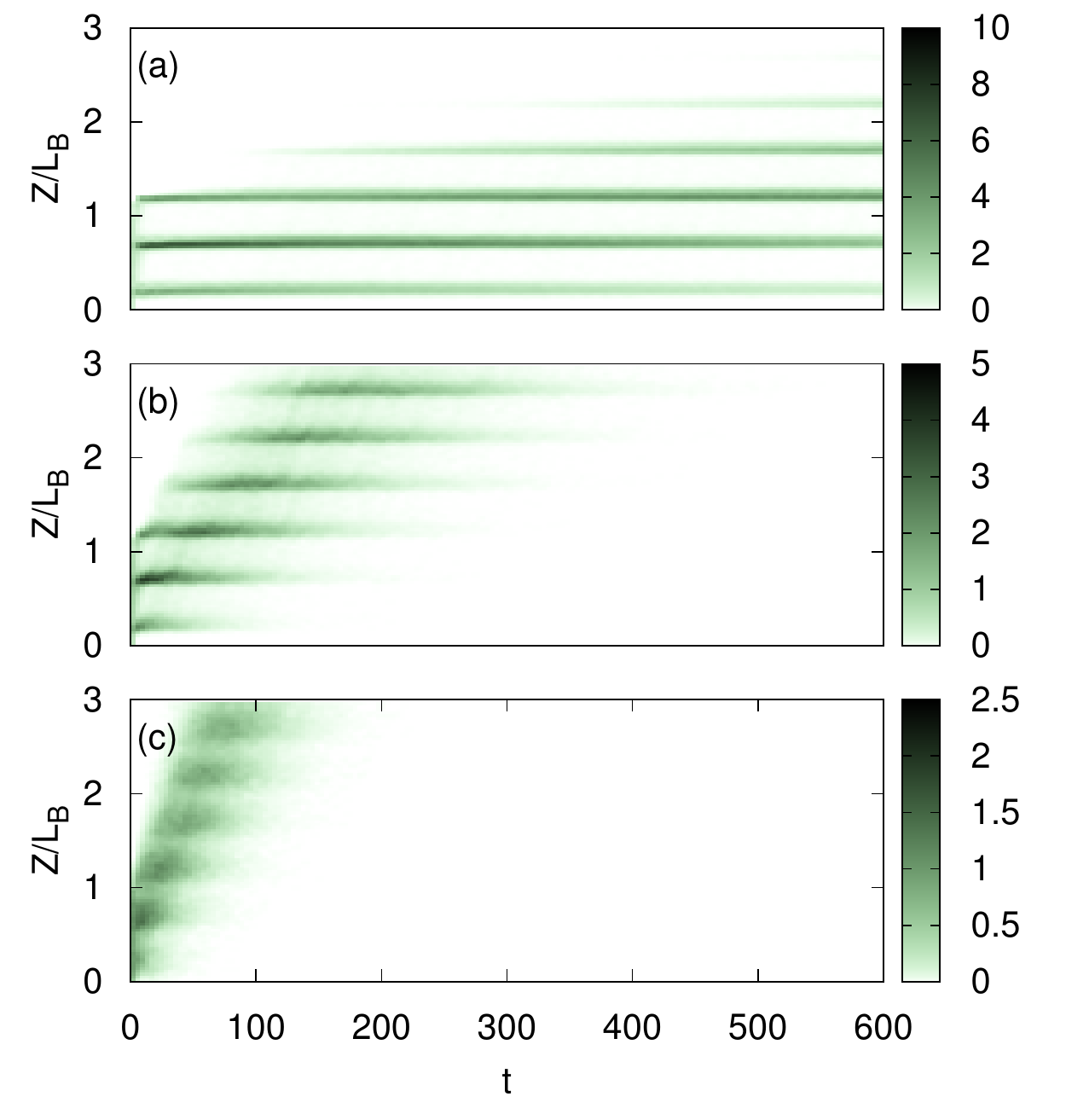}
\caption{Evolution of vertical density of cells in a turbulent
  Kolmogorov flow for swimming parameters $\Psi=1.1$, $\Phi=0.05$ and
  turbulence intensity $\gamma=0.01$ (a), $0.05$ (b), $0.2$ (c).  The
  initial condition is a random uniform distribution on $[0,L_B]$.
  Subsequent formation and disruption of layers is evident in panels
  (b,c) (compare with the laminar case in presence of rotational
  diffusion shown in Fig.~\ref{fig:densrot}). Time and scale have been
  made non-dimensional as for the laminar case.}
\label{fig:densturb}
\end{figure}

\begin{figure}[t]
\centering
\includegraphics[width=0.5\textwidth]{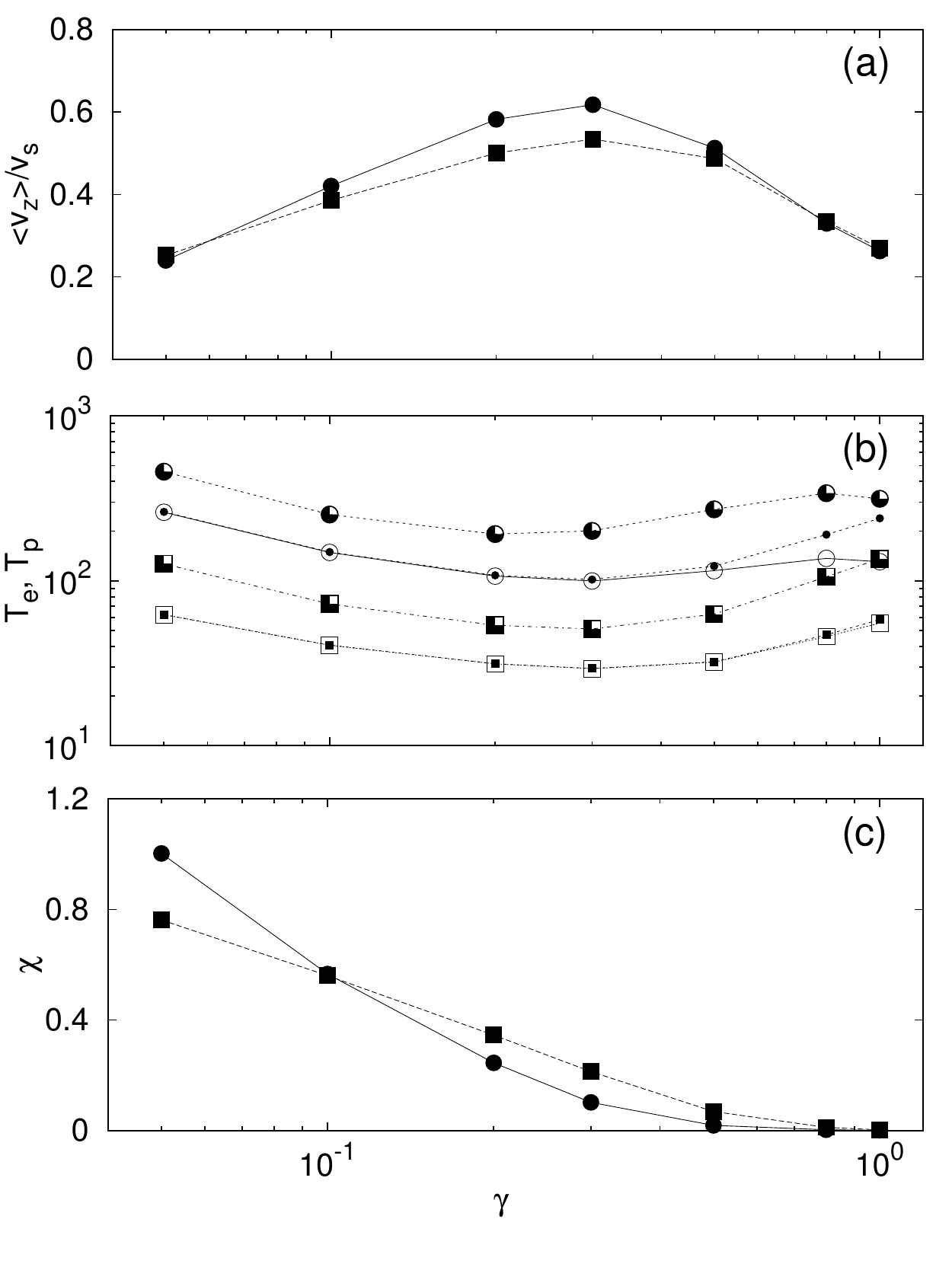}
\caption{Quantitative characterization of single cell and layer
  properties as a function of turbulent intensity $\gamma$, for
  $\Phi=0.05$ (circles) and $\Phi=0.2$ (squares) with $\Psi=1.1$. (a)
  Average vertical velocity $\langle v_z\rangle$ normalized to the
  swimming speed $v_s$.  (b) Layer persistence time $T_p$ (semi-filled
  symbols) and average trapping time $T_e$ (open symbols) for single
  trajectories, compared to $L_B/(2\langle v_z\rangle)$ (small filled
  symbols inside the empty ones). Notice the discrepancy between
  $\pi/\langle v_z\rangle$ and $T_e$ at $\Phi=0.2$, see text for a
  discussion. (c) Inhomogeneity index $\chi$ as from
  Eq.~(\ref{eq:chi}).
\label{fig:medie-turb}}
\end{figure}

Figure~\ref{fig:densturb} shows the formation and disruption of
layers, similarly to the phenomenology induced by RBM
(Fig.~\ref{fig:densrot}). More quantitatively,
Figure~\ref{fig:medie-turb} shows the average vertical velocity
$\langle v_z\rangle$, escape and persistence times $T_{e,p}$ and
inhomogeneity index as a function of $\gamma$, which plays a similar
role of $Pe_{r}^{-1}$ for the laminar case with RBM (compare with
Fig.~\ref{fig:medie}).


At moderate values of turbulent intensity, velocity and vorticity
fluctuations allow cells to escape from the trapping regions by moving
them to regions of lower shear, where upward directed swimming is
possible.  As a result, the average vertical cell velocity, $\langle
v_z \rangle$, which was zero in the absence of turbulent fluctuations,
becomes positive. However, very intense turbulence rotates the cell
swimming direction randomly and, moreover, fluctuations of the
vertical velocity also mix cells. As a consequence, the average
vertical motion $\langle v_z \rangle$ decreases for large values of
$\gamma$.  An intermediate turbulence intensity maximizes the vertical
migration velocity (see Fig.~\ref{fig:medie-turb}a).  The average exit
time, as already discussed, is determined by the average swimming
speed, i.e. $T_e =L_B/(2\langle v_z\rangle)$
(Fig.~\ref{fig:medie-turb}b).  However, unlike the laminar case
(Fig.~\ref{fig:medie}b), here the agreement is not perfect especially
when $T_e$ is large. The reason is that the total integration time of
turbulent simulations was shorter and the exit-time statistics does
not converge for very large $T$.  In realistic situations, where the
numerical or experimental time is finite, the evaluation of $T_e$ must
be carefully performed and compared with the other statistics such as
the average vertical velocity.  The most affected cases are the ones
with very weak or very intense turbulence. In the latter case
$\chi\approx 0$ and thus layering is negligible.

\begin{figure}[t]
\centering
\includegraphics[width=.45\textwidth]{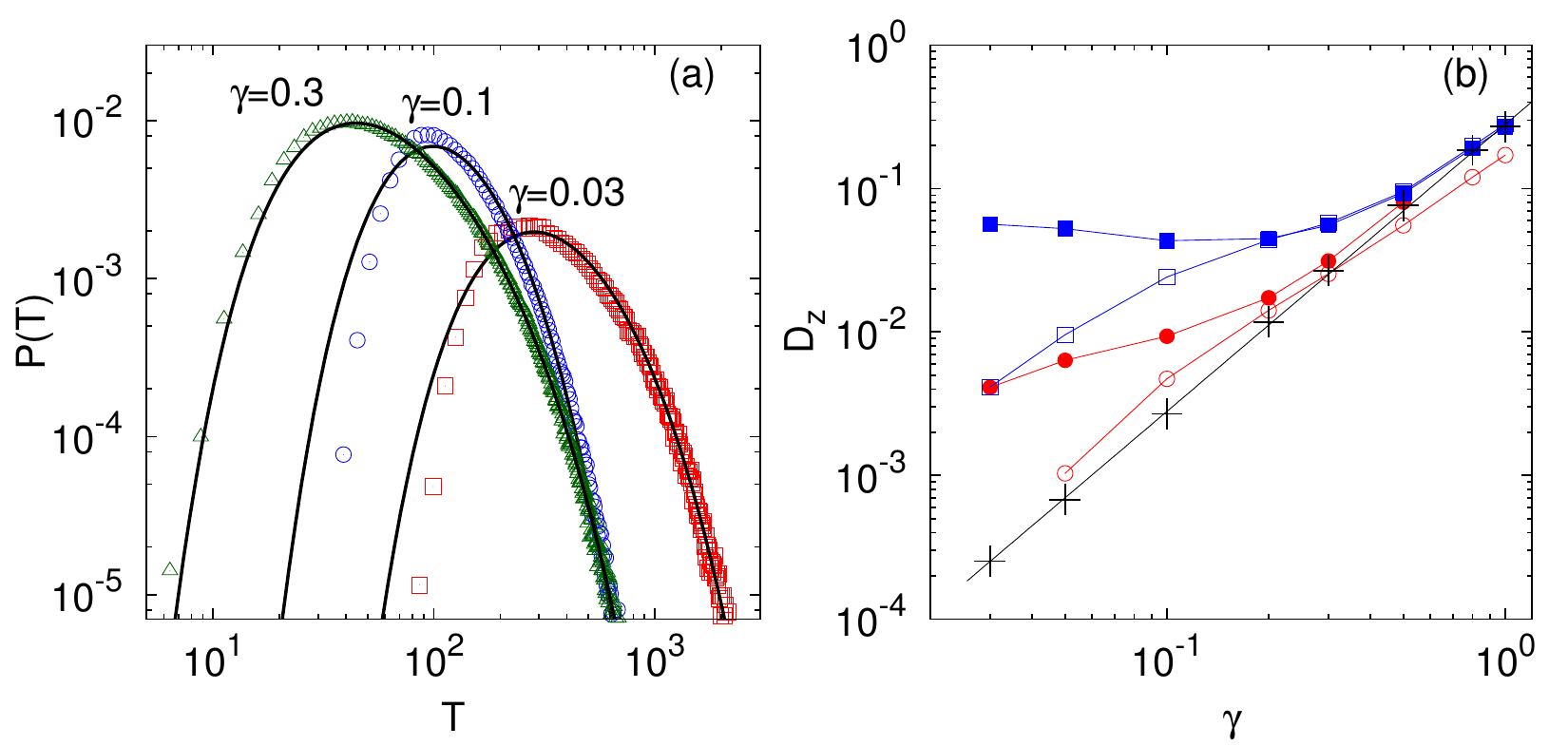}
\caption{(Color online) Exit time statistics. (a) Exit time PDF for
  $\Psi=1.1$, and $\Phi=0.05$ at three turbulent intensities $\gamma$
  as labeled, compared with the prediction (\ref{eq:ig}) with $D_z$
  and $V_d$ obtained as explained in the text. (b) Vertical
  diffusivity constant $D_z$ estimated from exit time statistics for
  cells with $\Phi=0.05$ (red circles) and $\Phi=0.2$ (blue squares)
  for $\Psi=1.1$ (filled symbols) and $\Psi=1.5$ (empty
  symbols). Statistical convergence is poor at large and small
  $\gamma$'s for $\Phi=0.05$.  The solid black line displays the
  $\gamma^2$ behavior of the turbulent diffusivity measured along
  tracer trajectories (plus symbols).
\label{fig:IG}}
\end{figure}

A simple way to model the vertical dynamics of gyrotactic swimmers is
in terms of a diffusive process with drift $V_d$ (due to the average
vertical migration speed, i.e. $V_d=\langle v_z\rangle$) and diffusion
constant $D_z$, whose value depends on turbulent fluctuations. The
escape from a layer can then be addressed through the exit-time
statistics asking for the time $T$ needed for a swimmer to travel a
distance $L_B/2$.  This is a standard problem in stochastic processes,
see e.g. Ref.~\onlinecite{Redner}. In the case of diffusion with
drift, the probability density function of the exit time $T$ is given
by the so called inverse Gaussian function, which we can write as
follows
\begin{equation}
\mathcal{P}(T)= \frac{L_B}{(4 \pi D_z T^3)^{1/2}} e^{-\frac{(V_d T -
    L_B/2)^2}{4D_z T}} \,.
\label{eq:ig}
\end{equation}
We thus have a prediction for the exit time PDF which can be directly
tested against the measured one.  For the drift velocity we have
$V_d=\langle v_z \rangle$, which is measured in DNS
(Fig.~\ref{fig:medie-turb}a), notice also that Eq.~(\ref{eq:ig})
implies $T_e= \langle T \rangle= L_B/(2 V_d)$, consistently with
Fig.~\ref{fig:medie-turb}b. The diffusion constant $D_z$ can be estimated by
measuring $\langle T^2 \rangle$ in the DNS and noticing that in
Eq.~(\ref{eq:ig}) $\langle T^2\rangle=L_B(D_z+LV_d/2)/V_d^3$.
In Figure~\ref{fig:IG}a we show the comparison between measured exit-time PDF
$p(T)$ and the inverse Gaussian prediction (\ref{eq:ig}), with $D_z$ and $V_d$
obtained as discussed above. The prediction turns out to be very accurate for
the right tail (long exit times) for all turbulent intensities $\gamma$, while
the left tail reproduces the numerical results only for  large values of
$\gamma$.  Indeed a purely Gaussian model cannot be expected to describe the
escape-time statistics accurately in presence of strong trapping. The
deviations in the left tails can be interpreted as the result of the
suppression of fast escapes due to gyrotactic trapping, which is more effective
in the limit $\gamma \to 0$.  On the other hand, long escape times allow
trajectories to sum-up many uncorrelated contributions, thus recovering a
diffusive behavior, which explains the good agreement on the right tail.  

It is natural to identify $D_z$ with the vertical
  turbulent eddy diffusivity $D_z^{turb}$, characterizing the large
  scale diffusive properties of fluid tracers.  In
Figure~\ref{fig:IG}b we show, as a function of the turbulent intensity
$\gamma$, the $D_z$ estimated from the exit-time statistics for
different values of $\Phi$ and $\Psi$ and the (vertical) turbulent
diffusivity measured from the vertical mean square displacement of
tracer particles in the same flow.  We start by noticing that the
turbulent diffusivity behaves as $D^{turb}_z \propto \gamma^2$, which
is consistent with the expectations as $D^{turb}_z \propto (u_{\rm
  rms}^\prime)^2 \approx \gamma^2 (U/2)^2$.  Then we observe that the
diffusivity $D_z$ for gyrotactic swimmers, estimated from exit times,
is typically larger than the turbulent one, and the deviation is more
pronounced at small $\gamma$ for larger swimming number and smaller
stability number, more precisely for $\Psi$ closer to $\Psi_c$. These
features can be rationalized as follows. In the presence of turbulent
fluctuations the vertical diffusivity is expected to have two
contributions: one from the fluctuations of the vertical velocity,
which can be estimated from turbulent diffusivity, and one from
swimming combined with the reorientation of the swimming direction due
to vorticity fluctuations.  Clearly, the latter contribution will lead
to a diffusivity which increases with the swimming speed and thus is
more important than the former for small $\gamma$, indeed $v_s/u_{\rm
  rms}^\prime \sim \Phi/\gamma$.  This explains the larger discrepancy
at small $\gamma$. As for the effect of stability, diffusivity due to
swimming is expected to be larger when vertical motion is more
coherent, i.e. cells are more stable in their orientation (i.e.
$\Psi$ is smaller). Finally, when turbulent fluctuations become the
dominant effect, layers tend to disappear and cells are expected to
recover a diffusive dynamics which explains the convergence of $D_z$
to the turbulent value for $\gamma \to 1$. Consistently, the PDF of
exit-times converges to (\ref{eq:ig}) (see Fig.~\ref{fig:IG}).

\section{Conclusions} 
\label{sec:6}

In this paper we have investigated the phenomenon of gyrotactic
trapping which has been recently proposed
as one of the possible mechanisms responsible for thin
phytoplankton layer formation.\cite{Durham2009,Durham2012}

We derived a detailed theory of the mechanism within the
framework of dynamical systems theory for the laminar Kolmogorov flow.
The  ideas and tools here developed can be generalized to basically any
laminar shear flow. In particular, the approach developed in
Sect.~\ref{sec:3} can be easily extended to
Poiseuille-like velocity fields such as that used in the experiments
presented in Ref.~\onlinecite{Durham2009}. Nonetheless the Kolmogorov flow
is advantageous as it allows us to avoid considering boundaries and
focus on bulk properties without the need to model the behavior of
swimmers close to walls. Moreover, thanks to the fact that the
sinusoidal mean profile is preserved in the turbulent case,
we studied how gyrotactic trapping is
altered by turbulence.

We found that turbulent fluctuations, similarly to random fluctuations
of swimming direction due to rotational Brownian motion, make
gyrotactic trapping transient. We characterized the phenomenon in
terms of trapping (or exit) times and showed that in the presence of
turbulence the statistics of exit times can be modeled (at least for
long trapping events) by a diffusive process with drift. In
particular, the (vertical) drift velocity results from the average
upward swimming while diffusion results from both turbulent
diffusivity (as for tracer particles) and swimming combined to
fluctuations of the swimming direction.  When velocity fluctuations
are small compared to the swimming speed, the diffusivity induced by
the latter is important.  As a consequence, care should be taken when
estimating the effect of turbulence on thin layers formed by swimming
phytoplankton in terms of turbulent diffusivity alone.  Swimming
combined to reorientation of the swimming direction can indeed be very
important for the diffusivity properties, as also recently recognized
in simple linear flows.\cite{lauga2014}

We found that the average exit time of single trajectories $T_e$ is
given by the average swimming speed and typical vertical length
characterizing the layers (which in the periodic Kolmogorov flow is
$L_B/2$). The persistence time of the layer $T_p$ is of the order of a
few (typically $\sim 2-3$) $T_e$ depending on the value of the
vertical diffusivity, using e.g. the inverse Gaussian prediction
(\ref{eq:ig}). Thus ignoring the aforementioned difficulties in
estimating $D_z$ for realistic oceanic flows, if we consider average
swimming speed $\langle v_z \rangle$ in the range $0.2-0.6 \, v_s$, as
suggested by Fig.~\ref{fig:medie-turb}b (with $v_s\approx 100-200 \mu
m/s$) and typical lengths $L_B$ of the order of a few centimeters, we
obtain an estimation of $T_p$ from a few hours to a few days, which is
akin to values found in TPLs observed on the field.

We conclude mentioning that it would be very interesting in future
investigations to consider a more realistic model in which modulation
of turbulent fluctuations are controlled by stratification, as in real
oceans. Moreover, it would be useful to quantitatively assess the
diffusion properties of swimming microorganism under the combined
effect of swimming and fluid motion.

\begin{acknowledgments}
We acknowledge useful discussions with Roman Stocker and Roberto
Tateo. 
\end{acknowledgments}

%

\end{document}